\documentclass{article}

\usepackage[pdftex,colorlinks=true,plainpages=false,pdfpagelabels]{hyperref}
\usepackage{latexsym}
\usepackage{amsmath,amsfonts,amstext,amssymb,amsbsy,amsopn,amsthm,eucal}
\usepackage{graphicx,bm,epstopdf}
\def\beq{\begin{eqnarray}}
\def\eeq{\end{eqnarray}}
\def\bea{\begin{eqnarray*}}
\def\eea{\end{eqnarray*}}

\def\centeron#1#2{{\setbox0=\hbox{#1}\setbox1=\hbox{#2}\ifdim
\wd1>\wd0\kern.5\wd1\kern-.5\wd0\fi
\copy0\kern-.5\wd0\kern-.5\wd1\copy1\ifdim\wd0>\wd1
\kern.5\wd0\kern-.5\wd1\fi}}
\def\ltap{\;\centeron{\raise.35ex\hbox{$<$}}{\lower.65ex\hbox{$\sim$}}\;}
\def\gtap{\;\centeron{\raise.35ex\hbox{$>$}}{\lower.65ex\hbox{$\sim$}}\;}

\def\singleandthirdspaced{\baselineskip=\normalbaselineskip\multiply
    \baselineskip by 130\divide\baselineskip by 100}

\newcommand{\newc}{\newcommand}
\newc{\qbar}{{\overline q}}
\newc{\Kahler}{K\"ahler }
\newc{\deltaGS}{\delta_{\rm GS}}
\title{  
\vspace*{-2.3cm}  
\begin{flushright}  
\normalsize{  
SCIPP-12/10
}  
\end{flushright}  
\vspace{1.5cm}  
\Large  
\textbf{
Deformed Bubbles and Lorentz Invariance in Vacuum Decay
\\
}\vspace*{1.0cm}   
}

\author{Michael Dine, Patrick Draper, and Chang-Soon Park
\vspace{5mm}
\\ 
\normalsize\emph{Santa Cruz Institute for Particle Physics and Department of Physics, Santa Cruz CA 95064} 
}

\date{}

\begin{document}
\setcounter{page}{0}  
\maketitle  

\vspace*{1cm}  
\begin{abstract}
Recently, questions have been raised about the role of Lorentz invariance in false vacuum decay.  It has been argued that infinities may arise in an integration over Lorentz-boosted final states. This suggestion motivates a Minkowski-space analysis of the decay rate.  
We attempt to illuminate features of the amplitude computation, and argue that the total rate including excitations is both finite and Lorentz invariant.
\end{abstract}

\thispagestyle{empty}  
\newpage

\vspace{-3cm}

\setcounter{footnote}{0} \setcounter{page}{2}
\setcounter{section}{0} \setcounter{subsection}{0}
\setcounter{subsubsection}{0}

%%%%%%%%%%%%%%%%%%%%%%%%%%%%%%%%%%%%%%%%%%%
%%%%%%%%%%%%%%%%%%%%%%%%%%%%
\singleandthirdspaced

\section{Introduction:  The Prospect of Infinities in Vacuum Decay}

The problem of vacuum decay, both in the presence of gravity and without, has been studied for almost four decades.  A general theory
of vacuum decay in systems with large numbers of degrees of freedom was developed in~\cite{banksbenderwu,banksbender}.  Early efforts to calculate the false vacuum decay rate in field theory from a Minkowski space/WKB
perspective appeared in~\cite{okun}.  These calculations provided a rough picture, but they raised certain questions, including a question of finiteness; it was argued that it is necessary to integrate over Lorentz transformations of the basic process, leading to a factor of the volume of the Lorentz group in the total rate.
The critical step was taken by Coleman (and Callan and Coleman)~\cite{coleman,callancoleman}, who formulated the problem at weak
coupling in the language of the Euclidean path integral treated
in the semiclassical approximation.  The tunneling rate was given by the exponential of a large, semiclassical term (the WKB term) times a prefactor, obtained from fluctuations about a solution of the Euclidean equations of motion.  The question of finiteness, it was argued, was solved; indeed, all infinities appearing in the calculation of the prefactor
could be absorbed in standard field theory counterterms.
In later work, Vilenkin and Vachashpati discussed the problem of constructing a wave functional describing vacuum decay~\cite{vilenkin}.  They argued that their
resulting construction was Lorentz invariant.

Coleman and De Luccia subsequently developed the theory of vacuum decay in general relativity~\cite{colemandeluccia}.
One of their remarkable observations concerned the decay of Minkowski space to a negative cosmological constant configuration; they found that such decays either do not
occur, or lead to gravitational collapse.  Most subsequent researchers  have assumed that this indicates that the fate of Minkowski  space, at least
in the cases of apparent collapse, cannot be investigated within quantum field theory (see, e.g.,~\cite{banksdecay}).  An interesting attempt to model the process taking advantage of AdS/CFT duality appears in~\cite{horowitz,eliezer}.

Recently, Dvali~\cite{dvali} has revisited these questions and argued that decay of a false vacuum in field theory on Minkowski space does not make sense.  Subsequently, 
Ref.~\cite{dvali2} delineated theories which might be discarded based on these considerations.  

The basic argument, like the old discussion of~\cite{okun},
has nothing to do with general relativity, but instead represents a new version of the argument that one is required to sum over Lorentz transformations of some decay
configuration.  Ref~\cite{dvali} notes that if a state can decay to a lower energy state through bubble formation, it can decay to a slightly larger bubble, with larger negative energy, and some
positive energy particle.  Such a configuration will not be Lorentz invariant, and it is argued that the total rate will be infinite after integrating over all Lorentz transformations
of the configuration.

One objection to this argument immediately comes to mind.  Because the false vacuum is invariant under spatial translations but not time translations (it decays), it cannot be invariant under Lorentz boosts. Therefore the rate to nucleate a bubble along with a particle or excitation with classical total four-momenta $(\pm E, \pm p)$ is not the same as the rate to produce a boosted bubble plus excitation with appropriately boosted four-momenta. This fact was mentioned in a footnote in~\cite{dvali}, but not elaborated\footnote{Ref.~\cite{dvali} speaks also of nucleating two bubbles, with momenta $(\pm m,0)$.  This problem is of higher order in the
semiclassical analysis, and requires an analysis distinctly more involved than we will consider in this paper, but, as we will see, there is no reason
to expect an infinity in this case either.}.  The issue can also be formulated in position space (or most precisely by considering the particle as a localized wave packet.)  There we expect that the amplitude for production of the bubble and particle depends on multiple length scales, including the relative coordinate, the size of the bubble, and the lifetime: for example, if the particle is produced kiloparsecs away, one would expect the amplitude to be highly suppressed.  Of course, these general arguments do not automatically preclude a divergence, only one proportional to the volume of the Lorentz group. Therefore we are motivated to study the production of excitations more carefully.

As is well-known, the calculation of first quantum corrections to the imaginary part of the false vacuum energy was first done by Callan and Coleman in the Euclidean formalism~\cite{callancoleman}.
In the Euclidean computation, there is no signal of such an infinity at any order of
perturbation theory about the bounce.  Indeed, Coleman proved~\cite{colemanmartin} that the lowest action Euclidean
configuration is always $O(4)$ symmetric.  As a result, there is no need (at least for these bubbles) on the Euclidean side even for an integration over $O(4)$.  On the Minkowski side, this translates
into a classically $O(3,1)$ (i.e. Lorentz invariant) nucleated state and its perturbations.  
If other non-$O(4)$ symmetric bounce solutions exist and were to be included in the computation, because of the compactness of $O(4)$, the Euclidean computation would remain finite. 
For there to be an infinity,
the Euclidean calculation would have to fail to capture the Minkowski space physics in a drastic way.

Although the Euclidean computation is clean and straightforward, we can attempt to see what happens in Minkowski space directly by summing over the rates to nucleate all excitations about some basic bubble configuration.  
It would be very surprising if the Minkowski and Euclidean calculations differed in the final result. Certainly an infinite result would involve a loss of unitarity, or a failure of the continuation from Minkowski to Euclidean space.

In an infinite Minkowski universe, in order to make sense of the false vacuum, we need to think of it as the limit of a sequence of long-lived states in which the field expectation value in increasingly large spatial regions lies at the local minimum of the potential. We can imagine preparing the universe in a ``false vacuum" state from this sequence by heating and cooling the universe so that it transitions thermally to this state. We will assume that such `boundary' issues do not have a strong impact on the lifetime of the approximate false vacuum state, and
take as our starting point Coleman's
original description of the Minkowski space tunneling probability.  He noted that the analytic continuation of the $O(4)$ symmetric 
bounce solution describes the nucleation of a perfectly
spherical bubble at a particular
space-time point (in Minkowski space).  One can consider the nucleation of the bubble, along with an infinite possible set of excitations.   We will focus principally on the thin wall approximation, which already incorporate the basic issues.  We will describe a suitable set of basis states for
the production of these excitations upon nucleation.  The treatment of low energy excitations, as we will see, is complicated.
The notion of the nucleation time, for example, is not extremely sharp\footnote{To quote Coleman, ``Like all descriptions of quantum-mechanical processes in the language of the old quantum theory, this one must be taken with a large grain of salt; it will certainly lead us astray if we try to use it to describe measurements made just outside the potential barrier.  Nevertheless, it is very useful as an asymptotic description."~\cite{colemanuses}}.  
But as our interest is in the role of highly excited states, this will not be an
obstacle.  

Our goals, then, will be to:
\begin{enumerate}
\item  Define the basis states for our computation of the nucleation of excitations of the spherical bubble
\item  Develop a methodology for computing the nucleation of these states and determine if the rate is finite
\item Understand the action of Lorentz transformations on the states and the total rate
\end{enumerate}

The rate to produce a symmetric bubble with an excitation has precisely the features of the particle/bubble pairs of Ref.~\cite{dvali}.  In particular, in order to produce a positive-energy excitation, it is necessary that the bubble, at nucleation,
be larger than the critical bubble (corresponding to an excitation of the $\ell=0$ angular mode, which is of negative energy.)  For the lowest lying excitations, the computation is challenging, but for the question
of infinities, it is enough to consider high energy excitations, and for these the analysis is reasonably straightforward.  We will see that, precisely because the system transits further in
the effective potential for the bubble radius, these high energy excitations are cut off exponentially, with a decay rate $\Gamma$ suppressed
by
\begin{align}
\Gamma \sim \exp\left({-\rm{a}\times[\Delta E]^{3/2}}\right)\; ,
\end{align}
where a is an ${\cal O}(1)$ constant and $\Delta E$ is the energy of the excitation.
 At a classical level, the action of Lorentz transformations map bubble states with average radius greater than that of the critical bubble to new bubbles with even larger radius.  Lorentz invariance of the sum is  immediate, as can be seen
by considering the problem from a ``passive" viewpoint.  Because the undisturbed bubble (no radial or positive energy excitations) is Lorentz invariant, it has a well-defined nucleation time in any frame, and takes the same
form in any frame; {\it as do its spectrum of fluctuations}.  As a result, the rate is the same in any frame.  From the active viewpoint, the states simply furnish a representation of the Lorentz group, and no further states are required. The total rate is then expected to be finite, whether integrating over boosts or summing over excitations.

Having understood the symmetric case,
we will then discuss $O(4)$ non-invariant solutions of the Euclidean equations with false vacuum-false vacuum boundary conditions at $t\rightarrow\pm\infty$.  To our knowledge,
no one has exhibited such solutions.  In fact, we will show that there is no $O(4)$ non-invariant solution
in any situation where the thin wall approximation is valid (to the best of our knowledge this is a new
result.)  On the other hand, there is no reason to think that, more generally,
such solutions don't exist.   For any would-be $O(4)$ non-invariant solution (which also must have an odd number of negative-eigenvalue fluctuations in order to contribute to the decay rate),  the puzzle is most stark.
On the Euclidean side, one must perform an $O(4)$ collective
coordinate integral, but this gives a finite factor of the volume of $O(4)$.  On the other hand, the Minkowski space continuation of the solution (if it exists) no longer possesses classical $O(3,1)$ invariance.  
An infinite set of distinct solutions can be constructed by acting with Lorentz transformations on a given solution, and we might think
that we must somehow ``integrate over all of these paths."  But, as in the $O(4)$ symmetric problem, we should be precise about what to compute.  If one considers the
probability of materialization of the asymmetric bubble on some space-like surface, the situation is complicated.  For a typical bubble, there is no time where the fields are static,
for example, and determining a suitable complete set of states is not simple, let alone computing the amplitudes to nucleate each. Absent a clearly formulated computation,
one cannot argue that there is a paradox. Indeed, as argued above in the limit where the bubbles are treated as particles, even making a correspondence between a family of classical Minkowski solutions and a complete set of quantum-mechanical final states, we would not expect to find equal tunneling rates. Furthermore, based on our experience with the symmetric
bubble, which already takes into account non-Lorentz invariant configurations (just at one higher order in perturbation theory), we see no reason to think that the Euclidean computation should fail.

The rest of this paper is organized as follows.
In sections \ref{setting} and \ref{leadingexponent}, we discuss the nature of the Minkowski problem as a WKB problem. (This subject has a long history, including
~\cite{okun,banksbenderwu,gervaissakita,vilenkin}).  We focus on the symmetric bounce, recalling how the leading tunneling exponent is obtained, as well as the
nature of the fluctuations.
In section \ref{adiabatic}, we discuss the production of excitations in an adiabatic
approximation.  We first explain how such corrections arise in a conventional Born-Oppenheimer bound state calculation, and then in the problem of decay.  We write explicit
formulae for the amplitudes for decay with excitations.  In section \ref{applications}, we apply the insights from this analysis to the vacuum decay problem.  We readily
see that amplitudes for highly excited states are exponentially suppressed, and that the overall rate is finite.  In section \ref{lorentz}, we explain why these results demonstrate
the Lorentz invariance of the final rate.  In section \ref{nonsymmetric}, we consider $O(4)$ non-invariant bounces. Finally, in an appendix, we comment on a second fashion in which large Lorentz boosts may alter the decay process by correlating multiple bubble nucleations. 

Our work has overlap with that of~\cite{vilenkindvali}.   These authors also note that the consequences of Lorentz invariance are not obvious.  But we differ in a number of details; e.g., they argue that ``a perturbative calculation of the probability for nucleating a negative energy bubble accompanied by a single positive energy excitation (`blob') is divergent" but say that this is an indication of the breakdown of perturbation
theory.  The difference arises, in part, because of our focus on space-like surfaces; indeed, they argue that it is integration over different times
that is the source of the problem.  They also assert that there are no $O(4)$ non-invariant bubbles, and thus do not consider the 
puzzle raised in this case. They do provide an illuminating, explicit example of a number of the issues through the Schwinger process.

\section{Setting of the Minkowski Problem}
\label{setting}

\begin{figure}
\begin{center}
\rotatebox{0}{\resizebox{70mm}{!}{\includegraphics{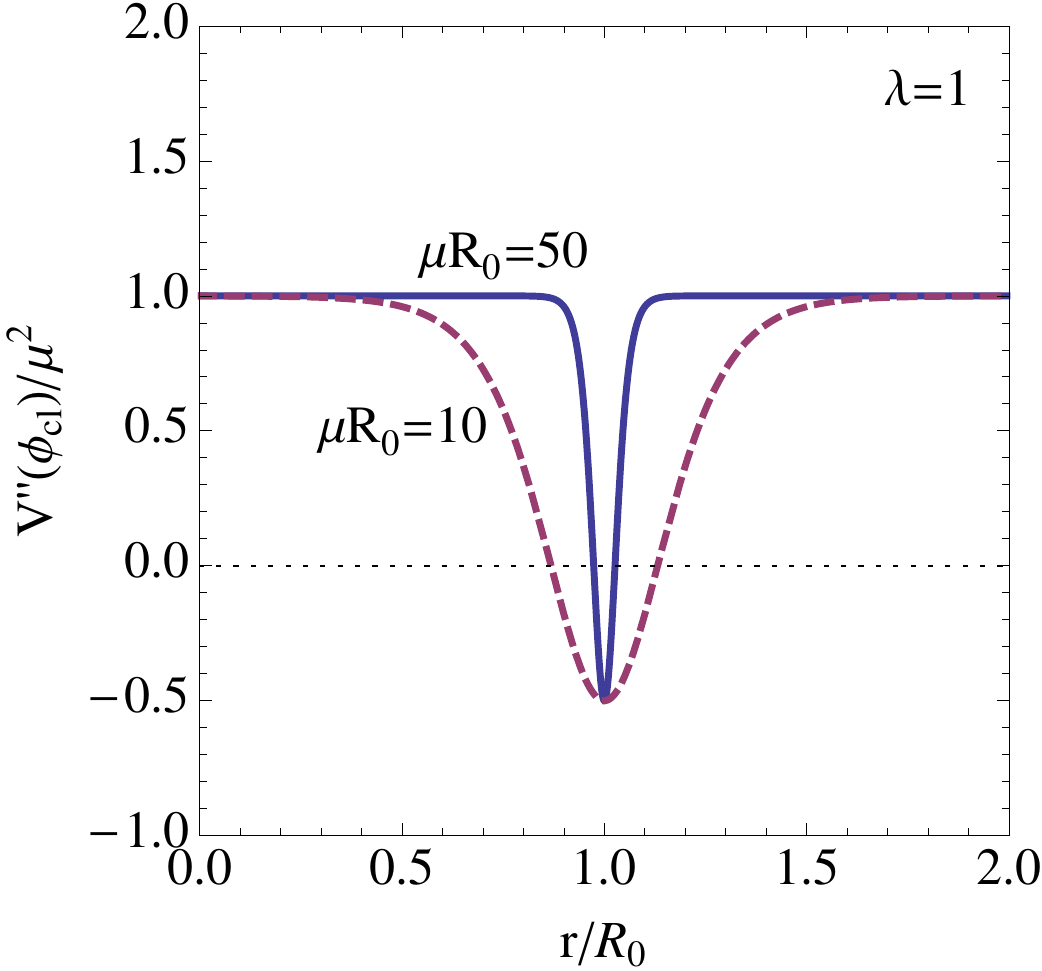}}}
\caption{Schematic form of the potential $U(r)$.}
\label{fig:Vpp}
\end{center}
\end{figure}

The argument of Ref.~\cite{dvali} applies perfectly well to the the thin wall limit, and the problem is most readily addressed there, as there is a natural separation of time and length scales.
Following Coleman, we will consider a theory of a single real scalar field, with a quartic potential in zeroth approximation, and a small splitting ($\epsilon$) between the two
degenerate states.  In the Euclidean calculation, one finds an $O(4)$ symmetric solution.  The bounce solution changes from false to true vacuum at $r  \sim R_0$,
 where the variation of the field occurs on a scale of order $\mu^{-1}$, the
inverse mass of the meson.
In the thin wall limit, $R_0$ is large, with size proportional to $1/\epsilon$.  The cross section of the wall is then that of a domain wall connecting the two,
approximately degenerate, vacua.
$S_1$ is the tension of this domain wall and is proportional to $\mu^3$, and in terms of it the critical radius is given by
\beq
R_0 = {3S_1 \over \epsilon}\;.
\label{eq:rcrit}
\eeq
The continuation of the Euclidean bubble $\phi(r^2 + t_E^2 - R_0^2)$ to Minkowski space is a Lorentz invariant evolution $\phi_c(r^2 - t^2-R_0^2)$.
The bubble is nucleated at $t=0$ with zero velocity ($\dot \phi = 0$).
At the beginning, the bubble moves quite slowly; it takes a time of order $R_0$ for it to double
in size. 
So it is natural to expand the classical solution around $t=0$:
\beq
\phi_{cl}(\vec x,t) = \phi_ {cl}(\vec x,0) + {t^2 \over 2}\ddot \phi_{cl}(\vec x,0) + \dots.
\eeq
The equation for small fluctuations of $\phi$ at early times is given by
\beq
{d^2 \over dt^2} \delta \phi -\vec \nabla^2 \delta \phi + U(r) \delta \phi = 0\;,
\label{eq:fluct}
\eeq
where
\beq
U(r) = V^{\prime \prime} (\phi_{cl}(r,t=0)) .
\eeq
$\delta \phi$ can be expanded in eigenmodes of the differential operator
\beq
H = -\vec \nabla^2 \delta \phi + U(r) \delta \phi
\eeq
Since the variation of $U$ occurs principally at $R_0$, one can treat this problem analogously to familiar problems
in molecular physics, separating variables and approximating the orbital angular momentum term by
${\ell (\ell + 1) \over R_0^2}$. Taking 
\beq 
\delta \phi = {\chi \over r},
\eeq
we can start with the eigenfunctions of the radial operator
\beq
\left (-{\partial^2 \over \partial r^2} + V^{\prime \prime}(\phi_{cl}) \right ) \chi_n = \omega_n^2 \chi_n.
\eeq
This operator has an eigenfunction of negative eigenvalue, 
\beq
\chi_0(r,\Omega) ={\partial \over \partial r} \phi_{cl}(r-R_0,t=0),\;\;\;\;\;\;\; \omega_0^2=-2/R_0^2.
\eeq
$\chi_0$ can be combined with the $\ell=1$ angular modes to make translations, which are zero modes of $H$. \noindent Higher modes of the radial operator will have eigenvalues of order the meson mass-squared and there is a continuum extending to $\infty$ (see Fig.~\ref{fig:Vpp} for the schematic form of $U(r)$.)

The modes of excitation can then be divided into two classes; those with frequencies comparable to $R_0^{-1}$ and those with frequencies large compared
to $R_0^{-1}$.  The slow modes (at $t=0$) are those associated with the approximate translation zero mode of the solution:
\beq
\phi(\vec x,t) = \phi_{cl}(r-R(t),0) +\sum_{\ell > 0,m} a_{0,\ell,m}(t) Y_{\ell m}(\Omega){1 \over r}\chi_0(r-R) + \sum_{n>0,\ell,m} a_{n,\ell,m}(t) Y_{\ell m}(\Omega){1 \over r} \chi_n(r-R) 
 \eeq
 $$~~~~+\sum_{\ell,m}\int d\omega^2 ~  a_{\omega,\ell,m}(t) Y_{\ell m}(\Omega){1 \over r}\chi_\omega(r-R)$$
(The $n = \ell = m=0$ mode is omitted from the sum; this is accounted for by the mode $R(t)$. The continuum eigenvalues are labeled by $\omega^2$.)

$R(t)$ is a collective coordinate which describes the overall growth of the bubble.  Near $t=0$, the Lagrangian for this mode is
\beq
L(R,\dot R) = {4 \pi \over 2}S_1 R^2 \dot R^2 - V(R)
\label{raction}
\eeq
where
\beq
V(R) =  4 \pi S_1 R^2 - {4 \over 3} \pi \epsilon R^3
\eeq
For $t \sim R$, the motion is relativistic, and this Lagrangian is inadequate.
We will return to the case of large (Minkowski or Euclidean) time, when the motion is relativistic, in section \ref{relativistic}.

At small times and for large $\ell$ ($\ell \gg 1$), the $a_{n\ell m}$ obey the equation
\beq
\ddot a_{n \ell m} + \omega_{n \ell m}^2 a_{n \ell m} = 0
\eeq
where
\beq
\omega_{n\ell m}^2 = \omega_n^2 + {\ell (\ell +1) \over R_0^2}.
\eeq
For $\omega_{n\ell m}^2 \gg ({1 \over R_0})^2$, the period of the motion is short compared to the time required for substantial changes in $R$, and the system can be treated adiabatically for short times.     These features of the classical problem suggest an adiabatic approach to high excitations in the quantum problem.  

Our interest is to consider tunneling with excitations of the various modes.  We expect significant suppression with $\ell,m$ and the size of the corresponding occupation
numbers.
To understand this problem we proceed in steps.
\begin{enumerate}
\item  We review how, for the dynamics of the mode described by $R(t)$, a conventional WKB treatment reproduces the leading contribution to
 the tunneling rate found by Coleman in the thin wall approximation.
\item   We estimate the rate for production of excited states in the decay.  For energies large compared to $1/R_0$, we argue that this problem can be treated in a a
Born-Oppenheimer approximation.  We discuss an analog in a quantum problem with multiple degrees of freedom.
\item  From the experience with the analog problem,
we obtain the dependence of the amplitudes on the energy of excitations in the field theory.  We demonstrate that the amplitudes die off exponentially for energies of order $\mu$ or larger, and that the
resulting tunneling amplitude is finite.
\item   
We consider the question of Lorentz transformations and invariance.  In a semiclassical approximation we can study the action of Lorentz transformations on bubbles larger than the critical radius (nucleated along with a positive energy particle) and see that the result is a new bubble configuration (along with a boosted particle.) The new bubble is larger, so the partial decay rate is suppressed compared to the rate into the unboosted state. Overall Lorentz invariance is maintained in the sum over all bubbles and excitations.

\end{enumerate} 

In the next sections, we consider these issues in turn.

\section{Leading Tunneling Exponent}
\label{leadingexponent}

For small $t$, it is easy to see that the solutions of the equations of motion obtained from Eq.~(\ref{raction}) sweep out the solutions of the full Euclidean field theory
problem.   The bubble is nucleated where the potential vanishes, i.e. at $R_0$ given in Eq.~(\ref{eq:rcrit}).
Expanding the potential about this point,
\beq
V \approx (R - R_0) V^\prime
\eeq
where
\beq
V^\prime = -S_1 R_0.
\eeq
So for small times ($t \ll R$),
\beq
R(t) =R_0 + {1 \over  2 R_0}t^2.
\eeq
Let's compare with the behavior of Coleman's bubble at nucleation.  This is
\beq
\phi_{cl}(\sqrt{r^2-t^2} -R_0) \approx \phi_{cl} (r  - {t^2 \over 2 R_0})\;.
\eeq
So for small times, we see that we describe the classical solution.  More generally, we require a relativistic generalization.
This is also the case for small fluctuations.

\subsection{Relativistic Treatment}
\label{relativistic}

The $n=0$ modes correspond to displacements of the bubble wall:
\beq
R(t,\Omega) = R(t) + \delta R(t,\Omega) = R(t) + \sum_{\ell m} a_{\ell m} Y_{\ell m}(\Omega).
\eeq
\beq
\delta \phi = {\partial \phi_{cl} \over \partial r} \sum_{\ell m} a_{\ell m} Y_{\ell m}(\Omega).
\eeq
An approximate treatment of wall evolution, in the classical case, has been developed in~\cite{freese}.  This approach allows a straightforward relativistic treatment.  For the equation of motion for $R$, one finds
\beq
\ddot R = {\epsilon \over S_1} (1-\dot R^2)^{3/2} - {2 \over R} (1 -\dot R^2).
\label{eq:Reqmot}
\eeq
For a zero energy bubble, this gives:

\beq
(1-\dot R^2)^{1/2} = {3 S_1 \over \epsilon R}.
\eeq
For small (Minkowski) times, this agrees with our earlier solution.  
The associated action is that of the spherical bubble.
For large times, the motion becomes relativistic.  For the Euclidean problem, $\vert \dot R \vert$ grows at small $R$, behaving as
\beq
\vert \dot R \vert = {3 S_1 \over R}.
\eeq

For the fluctuations, calling $\Delta_{\ell m} = a_{0 \ell m}$, one obtains the equation
\beq
\ddot \Delta_{\ell m} + \dot R \left (3 {\epsilon \over S_1} (1 - \dot R^2)^{1/2} - {4 \over R} \right ) \dot \Delta_{\ell m}
+ {(1 - \dot R^2)(\ell (\ell+1) -2) \over R^2} \Delta_{\ell m} = 0.
\eeq
For small times ($\dot R \approx 0$), this agrees with our earlier equations for $a$.  For large $\ell$, the system can be treated adiabatically.  This approximation certainly
breaks down for Minkowski times for which $\dot R \sim 1$; the first derivative terms in the equation of motion dominate.  But for Euclidean times, when $\dot R$ is large, the
second and third terms in the equation are comparable for small $\ell$, and the third term dominates for large $\ell$.  So one expects an adiabatic treatment to be reliable
for large $\ell$, and not too misleading for small $\ell$.  These statements will carry over to the quantum treatment below.

\section{Higher Excitations:  Adiabatic Approximation}
\label{adiabatic}

The discussion above is suggestive of an adiabatic treatment of the problem of production of excitations during
tunneling..
To gain insight into such processes, it will be helpful to first consider some examples from quantum mechanical
systems with small numbers of degrees of freedom.  As a warmup, we write a formula for the first corrections to the Born-Oppenheimer
approximation.  We will phrase the problem in the language of molecular physics, but the generalization to other systems is immediate.
In the molecular physics problem, one has nuclei of mass $M_i$, and coordinates $\vec R_i$ (which will abbreviate simply as $R$), as well as electrons with coordinates $\vec x_\alpha$ (which we will abbreviate simply as $x$)
and mass $m$.  One defines the zeroth order problem by solving for the electron wave functions for fixed nuclear positions, $\psi(\vec x_\alpha;R)$.  The corresponding
energy eigenvalues define an effective potential for the nuclei.  One then solves for the wave function of the nuclei, $\Psi(R)$; the full
wave function is a product:
\beq
\Phi_{Nn}(R, x) = \Psi_{n,N}(R) \psi_n( x;R).
\eeq
These solutions, of course, are not exact.  They are eigenfunctions of the Schrodinger operator only if one neglects terms ${\partial \psi_n \over \partial R}$
in the Schrodinger equation.
Thinking semiclassically, one is neglecting the speed of the nuclei compared to that of the electrons.  We can obtain the wave function to the next order
of approximation by writing, say, for the ground state:
\beq
\Phi_{0}(R,\vec x) = \Psi_0(R) \psi_0( x;R)+ \sum_{N,n \ne 0} c_{Nn}
 \Psi_N(R) \psi_n( x;R)\;.
\eeq
Plugging in the Schrodinger equation, and keeping terms with one derivative with respect to $R$ acting on $\psi_0$
yields
\beq
c_{Nn} = {1 \over E_{Nn} - E_{0}} \int dR dx \Phi^*_{ Nn}  {1 \over 2 M} {\partial \psi_0 \over \partial R}{\partial \Psi_0 \over \partial R}\;.
\eeq
Now consider a system with a particle described by a coordinate $R$ and a metastable minimum in the potential.  Suppose that $R$ couples to an oscillator, $x$, with
frequency $\omega(R)$, a slowly varying function of $R$, with $\omega(R\rightarrow 0) = \omega_1$ and $\omega(R \rightarrow \infty) = \omega_0$.
We want to ask the amplitude to tunnel from the lowest state in the well to states with excitations of the oscillator.  This situation
is analogous to the field theory problem -- by tunneling further along the $R$ potential, one can create oscillator excitations of higher energy (arbitrarily high energy
in the field theory problem).  On the other hand, we might expect that there is a price for these excitations; in the WKB approximation, one would expect
a larger exponent (action) for highly excited states, suppressing the amplitude.

\begin{figure*}
\begin{center}
\begin{tabular}{cc}
\includegraphics[width=0.45\textwidth]{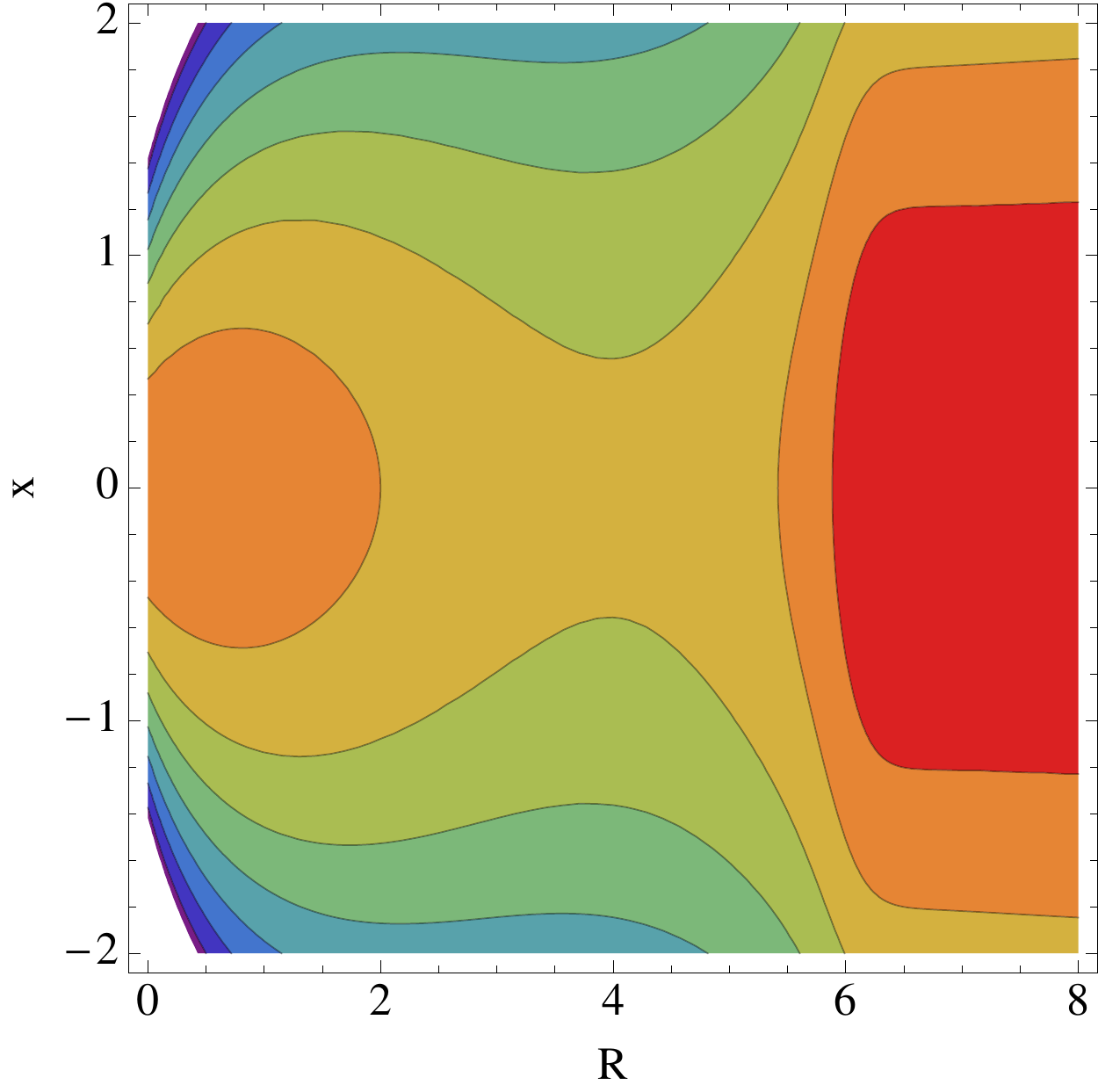} &
\includegraphics[width=0.45\textwidth]{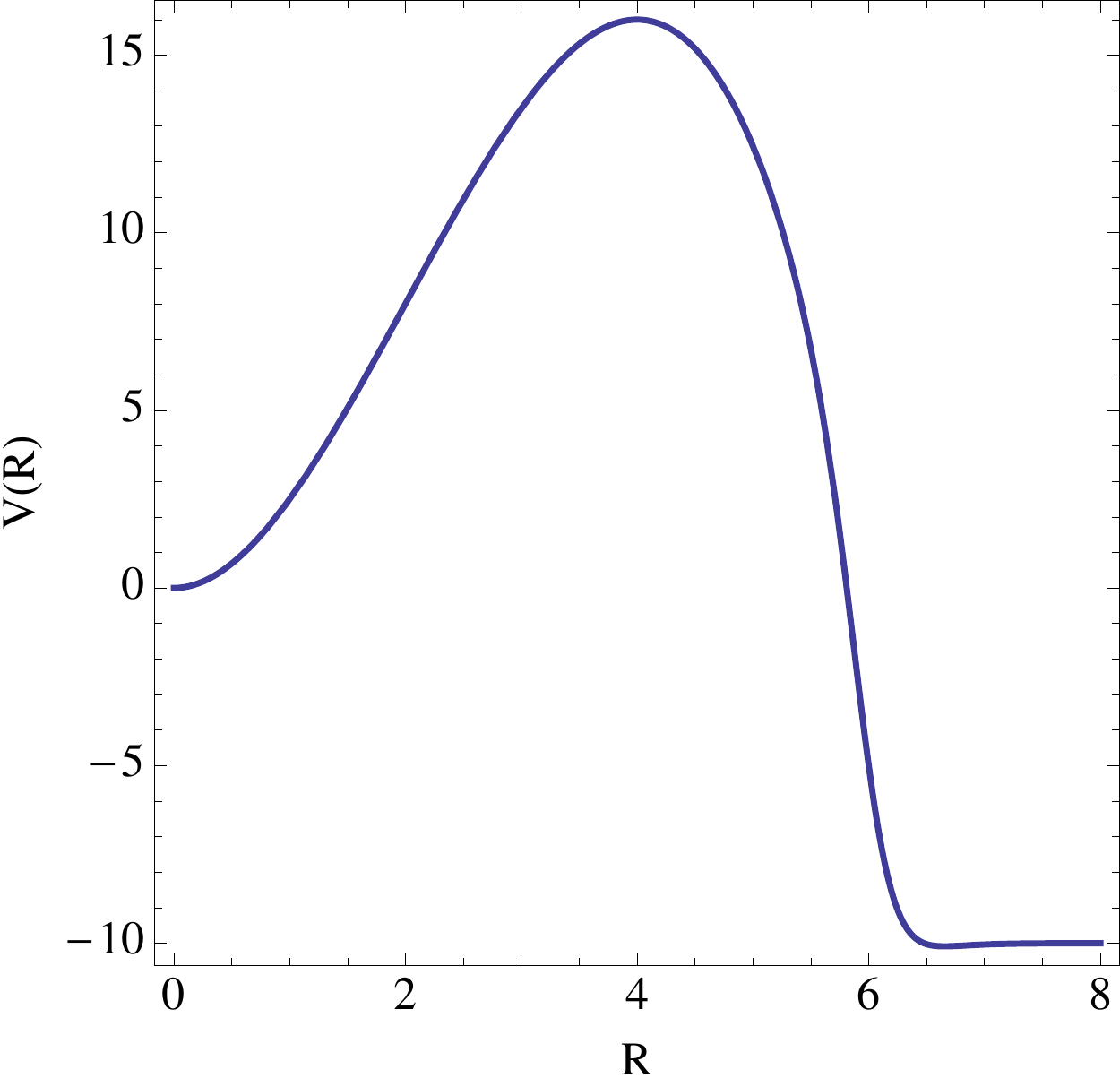} 
\end{tabular}
\caption{Left: A contour plot of a potential exhibiting a coupling between a harmonic oscillator $x$ and a coordinate $R$ with a metastable minimum (at the origin). For large $R$, the potential reaches a stable, flat value along $x=0$, and the harmonic oscillator decouples from $R$. Right: the $R$-potential for $x=0$.} 
\label{fig:potl}
\end{center}
\end{figure*}

To investigate this question, it is helpful to consider it as a scattering problem.  We take the potential to be
\beq
V(R,x) = V_0(R) + {1 \over 2} K(R) x^2
\eeq
where $V_0(0) = 0$, and $V_0(\infty) = -V_0.$    We suppose that ${\partial K \over \partial R}$ is everywhere small,
so that an adiabatic picture is appropriate, and that it goes rapidly to zero in the region where $V_0(R)<0$; this insures
that transitions between different oscillator states occur only under the barrier (this will not be the case in the field theory problem; we will discuss this issue in the next
section.) A sketch of the potential and its profile for $x=0$ are given in Fig.~\ref{fig:potl}.

We consider an initial state with total energy  $E_0$.
We can write the incoming wave function in the form:
\beq
\Phi(R,x)_{E_0,n} = \sum_{n=1}^\infty \Psi_{E_0,n}(R) \psi_n(x;R).
\label{productstate}
\eeq
Here $\psi_n(x;R)$ is the harmonic oscillator wave function appropriate to the $n^{\rm{th}}$
excitation with frequency $\omega(R)$, i.e.
\beq
\psi_n(x;R) \sim x_0^{-1/2}H_n(x/x_0) e^{-{x^2 \over 2 x_0^2}}
\eeq
where
\beq
x_0 = (\omega(R) m)^{-1/2}.
\eeq
To the right of the barrier,
$\Psi_{E_0,n}$ is the solution of the Schrodinger equation for $R$ with potential
\beq
V_n = V_0(R) + (n+ {1 \over 2}) \omega(R)
\eeq
with energy $E_0$, obtained in the WKB approximation.  Provided one can ignore ${\partial \omega \over \partial R}$,
$\Psi$ behaves, in the forbidden region, as
\beq
\Psi_{E_0,n}(R) \sim {\rm exp} \left (- \int_{R_n}^R dr \sqrt{{2 M} (V_n(r)-E_0) } \right )
\eeq
where $R_n$ is the turning point of the potential $V_n(R)$.
If the excitation number, $n$, is not too large, this can be estimated by studying the integration region near $R_0$.  Calling
$R_n = R_0 + \delta R$, 
\beq 
\delta R \approx {n \omega(R_0) \over V_0^\prime(R_0)}
\eeq
it is simple to estimate the additional contribution to the WKB exponent arising from the
integration region between $R_0$ and $R_0 + \delta R$.  For $R$ smaller than $R_0$, we can estimate the additional
correction to the WKB exponent by considering $R < R_0 - \Delta R$, $\Delta R \sim {\rm few} ~\delta R$), and  $n \omega \ll V_0$; then
one can expand the square root in powers of $\omega/V_0$ and again estimate the corrections to the WKB integral.  This yields the estimate:
\beq
\Psi_{E_0,n}(R) = \Psi_0(R) {\rm exp}\left (-n \int_{R_0-\Delta R}^R dr {1 \over 2}  {\sqrt{2M}\omega(r) \over \sqrt{V_0^\prime(R_0)\times(r-R_0)}}\right ) 
e^{-{c} \sqrt{2M} {1 \over V_0^\prime} [n \omega(R_0)]^{3/2}}
\label{nexp}
\eeq
where $c$ is a number of order unity, and this formula is valid  away from $R_0$.  We will estimate the size of these terms when we turn to the field theory.  But it should be noted at this stage that the second factor dominates, and quite generally, we expect that under the barrier, $R$ wavefunctions in states with oscillator excitations of energy $\Delta E$ will
 be suppressed by a factor
 \beq
 {\cal F} \sim  \ e^{-{c} \sqrt{2M} {1 \over V_0^\prime} [\Delta E]^{3/2}}
\label{energysuppression}
\eeq
In the regions in which ${\partial \omega \over \partial R}$ is not extremely
small, the $\Psi_{E_0,n}$  cannot be reliably computed in the WKB approximation.  The factor ${\cal F}$, however, still describes the leading
suppression of the amplitude for the particle to reach the barrier while still in the excited state, and tunnel through.

We are interested in resonances associated with the lowest state in the well (i.e. we want to take $E_0$ close to
this energy); the coefficient of the pole is proportional to the width of the state (the distance off of
the real axis is similarly proportional to the inverse lifetime).  At leading order in the Born-Oppenheimer approximation, the lowest state has overlap only with the ground state of the oscillator, and the usual analysis of the decay width corresponds to taking $n=0$ for the initial state.
However, we would expect that at higher orders in the approximation there will be overlap between the system's energy eigenstates and the leading-order product states with nonzero oscillator excitations. The decay rate into oscillator excitations will then receive perturbative contributions from overlap factors and exponentials of the form (\ref{energysuppression}).

\section{Application to the Field Theory Tunneling Problem}
\label{applications}

In the previous sections, we have considered tunneling in a system coupled to an oscillator.  We have seen that the final states can include not only the oscillator
in its ground state but in excited states as well.  Tunneling directly to these excited states is suppressed by the longer distance the system must tunnel through the forbidden region.
Our model was grossly simplified by the requirement that ${\partial K \over \partial R}$ vanish beyond the barrier.  In more realistic models, because of the exponential suppression of the
direct tunneling, the most favorable way to produce excited states in the far future is through tunneling to the low lying
 states, followed by ``jumps" to the higher levels.

The field theory problem possesses certain features which are different than those of our quantum mechanics example.  The potential, $V_0$, tends to $-\infty$ as $R \rightarrow \infty$, and
the frequencies of the low lying modes still vary in the allowed region.  As a result, after the nucleation of the bubble, excitations are continually produced.
Clearly to obtain the total rate for production of a given state at time $t$,
it is not sensible to simply sum over all of these different processes happening at different times.  Conceptually the simplest approach is to sum over all possible excitations at the moment of bubble nucleation.  The notion of ``moment of bubble nucleation" is not completely sharp, but is adequate for our considerations.

From the quantum mechanics example, we have seen that for states for which the energy is not too large, there is an exponential suppression
with energy (Eq.~(\ref{energysuppression})).  For very high energies, we would expect that the suppression is even stronger.  In the field theory
problem,
\beq
M = {1 \over 2} S_1 R^2\;;
~~~~~V(R) = 4 \pi S_1 R^2 - {4 \over 3} \pi \epsilon R^3.
\eeq
We will principally be concerned with the potential near $R= R_0$, so it is useful to note that
\beq
M \sim S_1 R_0^2;~~V^{\prime} \sim S_1 R_0;~~\omega \sim {\sqrt{\ell(\ell+1)} \over R_0}.
\eeq
Therefore, for large $\ell$, the suppression factor behaves as
\beq
 {\cal F}\sim \ e^{-{c}[\Delta E^3/S_1]^{1/2}}.
\label{fieldtheorysuppression}
\eeq
At any finite order of perturbation
theory about the bubble,  one does not expect exponential growth of the density of states, so we expect the overall rate to be finite.

\subsection{Low-Lying Excitations}

It is interesting to consider some of the types of excitations which might be nucleated along with the bubble.
In Section~\ref{setting}, we identified a complete set of states for the field theory at the moment of nucleation.  These were labeled by occupation numbers for levels $\omega,\ell,m$, where, up to terms of order $1/R_0^2$,
the $\omega$ were eigenvalues of the domain wall problem in two dimensions.  In addition to the zero eigenvalue, there
may be some number of discrete eigenvalues, and a continuum above that.  Consider first the discrete states at $n=0$.  
In comparison with the quantum mechanics problem, the frequencies are:
\beq
\omega_{\ell m} = {\sqrt{\ell (\ell+1)-2} \over R}.
\eeq
The $\ell=0$ mode corresponds to the collective coordinate $R$; the $\ell=1$ modes correspond to translational collective coordinates,
about which we will say more shortly.

Overall, one has a suppression of the total rate controlled by the sum: 
\beq
\sum_{\ell n_{\ell,m} } f(\ell)e^{-{ (n_{\ell,m} \ell )^{3/2} \over (S_1 R_0^3)^{1/2}}}. 
\eeq
where $f$ varies more slowly with $\ell$ than the exponential. This sum is convergent.  
The low-$\ell$ large-$n$ terms in the sum cannot be treated in the Born-Oppenheimer approximation, but we do not expect
a divergence from them.

\subsection{Particle Emission with the Bubble}
\label{particleboost}
Now consider the continuum.  Here we encounter directly the question of production of very high momentum/energy particles.  We again expect suppression, but we need to pay attention to questions of energy and momentum conservation. 

We first consider the problem from the point of view of the mode expansion.  Emission of a particle of energy-momentum $p^\mu = (E,\vec p)$ corresponds to emission of a state in the continuum; the energy-momentum
must be compensated by either the emission of another continuum state, or of one of the bound excitations.  Focusing
on the bound modes, in linearized approximation, and at 
$t=0$,
\beq
\vec P = \int d^3 x \partial_i \phi_{cl} \partial_0 \delta \phi + {\cal O}(\delta \phi^2).
\eeq 
This receives contributions only from the $\ell=1$ mode.
In this way, energy and momentum can be conserved in the emission of a high momentum particle by
\begin{enumerate}
\item
turning on a non-zero $a_{0,1,m} = {v t \over R_0}$, 
\item
varying the initial bubble radius,
\beq
R = R_0 + \Delta R
\eeq 
with
\beq
E = {\partial V \over \partial R_0} \Delta R.
\eeq
\end{enumerate}
Note, in particular, that the kinetic energy of the bubble (the $\int d^3 x {1 \over 2} (\partial_0 \phi)^2$ term in the energy)
is suppressed (it is of order ${P^2 \over S_1 R_0^2}$).
Momentum and energy can be conserved in other ways, for example, turning on other modes and working to higher order in
the fluctuation parameters, $a_{n,\ell,m}$.

For the excitation of modes with 
\beq
a_{0,1,0} = {v t \over R_0}; ~~\Delta R \ne 0\;,
\eeq
we have 
\beq
P_z = \int d^3 x \partial_z \phi_{cl} \partial_0 \delta \phi = \int d^3 x {x_i x_j v_j \over r^2} \psi_0^2(r-R)  
\eeq
$$~~~~={4 \pi v_i \over 3} R_0^2 \int dr \psi_0^2 = {4 \pi v_i \over 3} S_1 R_0^2\;,$$
\beq
E = {\partial V \over \partial R_0}\Delta R = - {4 \pi} S_1 R_0 \Delta R \;.
\eeq
It is straightforward to work out the Lorentz transformation of the fields through first  order in $\beta$.  For transformations along the
$z$ direction, there is a shift in the $\ell=1$ component and $\Delta R$:
\beq
\vec v \rightarrow \vec v -3 {\Delta R \over R_0}\vec  \beta;~~~~\Delta R \rightarrow \Delta R -\frac{1}{3} {\vec \beta \cdot \vec v} R_0,
\label{eq:galboost}
\eeq
corresponding to
\beq
E \rightarrow E + \beta P;~~~ P \rightarrow P + \beta E,
\eeq
as they must for a solution of the equations of motion. Note that for $\vec\beta$ antiparallel to $\vec v$, Eq.~\ref{eq:galboost} simplifies to $v\rightarrow v(1+\beta)$, $\Delta R\rightarrow \Delta R(1+\beta)$.
In addition, the boost generates an $\ell=2$ mode, but this contributes to the energy-momentum only at higher order in $\beta$.
At higher orders, one must be careful to work out the solutions to the equations of motion to higher orders in $t$ at small time.

\begin{figure*}[!t]
\begin{center}
\includegraphics[width=0.50\textwidth]{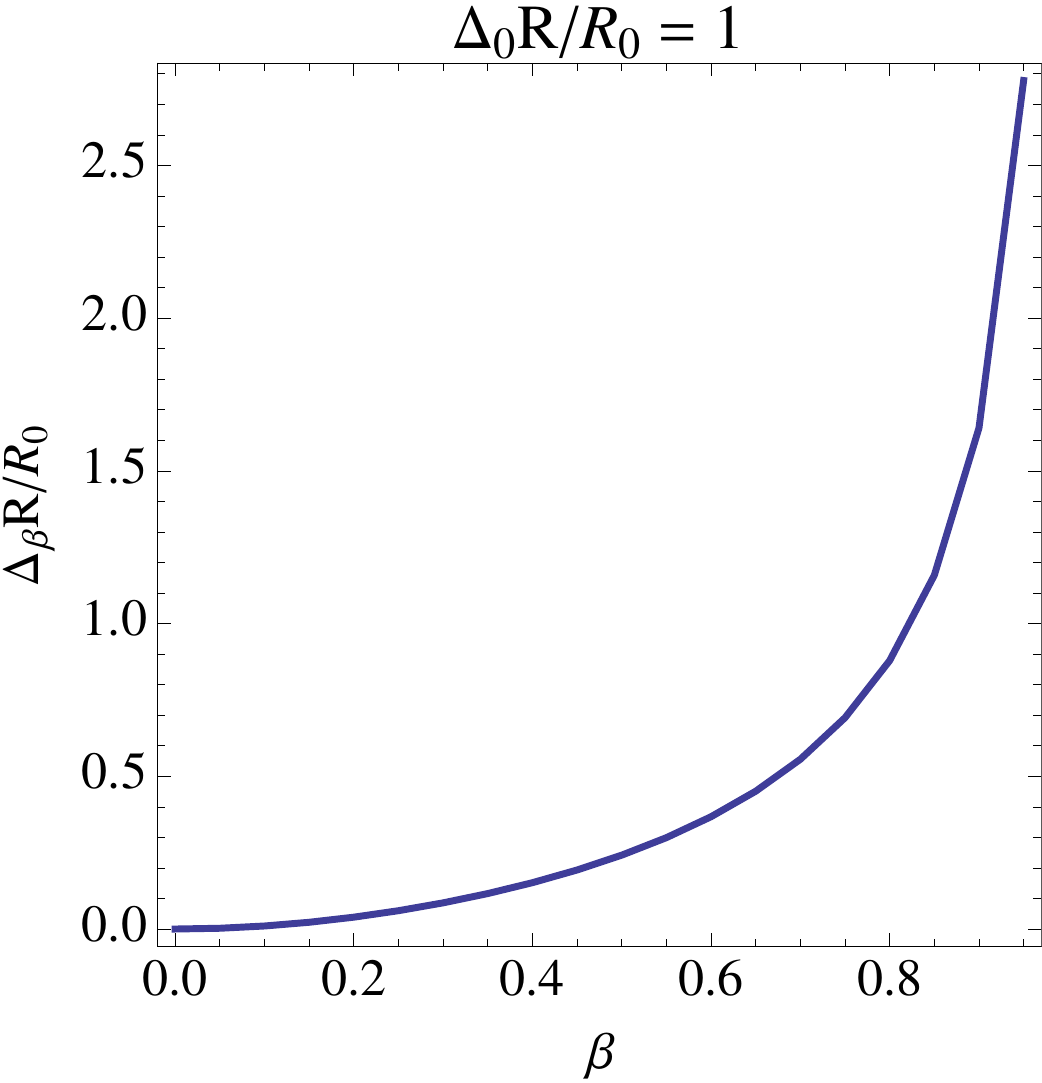} 
\caption{Growth of a bubble radius under Lorentz boosts of parameter $\beta$. The unboosted bubble is spherical and static with radius $R_0+\Delta_0R=2R_0$.}
\label{fig:largeboost}
\end{center}
\end{figure*}

Another way to produce a high momentum particle and conserve 4-momentum is to consider the nucleation of the spherical bubble and particle at rest, and apply a large Lorentz boost to the system. If the stationary bubble is larger than the critical bubble, it is not Lorentz invariant, and its rest energy/invariant mass should be the negative of the particle mass. In Ref~\cite{dvali}, it was pointed out that if the nucleation rate for the boosted system depends on the bubble-particle relative 4-momentum $p^\mu$ only as $p^2$, then an integration over boosts will diverge.  However, we expect that boosts will in general excite various modes of the bubble, which we can calculate in the classical limit using the classical time evolution of Eq.~\ref{eq:Reqmot}. In particular we expect boosts to lower the energy of the bubble further, which implies a growth of the bubble radius ($\ell=0$ mode) as we boost. Then, from our previous analysis, we expect exponential suppression of the rates. Indeed, as stated in the introduction, the breaking of Lorentz invariance in the rates is expected since the false vacuum initial state is not Lorentz invariant. 

When the particle is at rest, the bubble must satisfy (to leading order in $m/(S_1R_0^2)$)
\beq
\Delta_0 R \simeq {m \over 4 \pi S_1 R_0}\;,~~~\beta=0.
\eeq
Now define $R_i\equiv R_0+\Delta_0 R$ and consider Lorentz boosts of the system. Using the small-time equation of motion, valid to order $t^2$, we find that boosted bubble at $t=0$ satisfies to order $\beta^2$
\begin{align}
R_i^2-\left(1-3\beta^2\left(\frac{R_i}{R_0}-1\right)\right)x^2-y^2-z^2 =0\;.
\end{align}
Here we are using a membrane picture, which is equivalent to the field theory calculation above where we considered bound modes. The $\ell=0$ mode of the configuration has indeed grown, with 
\begin{align}
\Delta_\beta R \simeq \frac{1}{2} \beta^2 \Delta_0R\;,
\end{align}
 and the total $\Delta R = \Delta_0 R + \Delta_\beta R$. We see that at order $\beta^2$ there will be a contribution to the wavefunction suppression. To go to large boosts requires solving the full relativistic equation of motion for the spherical membrane, which can be done numerically. In Fig.~\ref{fig:largeboost} we plot the growth of $\Delta_\beta R$ as a function of $\beta$ for the case of an initial $\Delta R=R_0$, and observe that the growth of $R$ becomes rapid at large $\beta$. Note that in the numerical analysis $\Delta R$ does not need to be small. Eventually the linear approximation used in the rate suppression will break down, but since the potential falls in only a polynomial way, the exponential behavior of the suppression will continue.

\subsubsection{Suppression of the Tunneling Amplitude for Large Momentum}

Given that $\Delta R > 0$ for emission of a particle with non-zero momentum, there is a contribution to the suppression
of the tunneling amplitude of order

\beq
{\cal F} \sim e^{-c \left ({f(\vert \vec p \vert) \over S_1^{1/3} } \right)^{3/2} }\; ,
\eeq
where $f(\vert \vec p \vert)=\vert \vec p \vert$ if only $\ell=1$ is activated (as in the first example above), and $f(\vert \vec p \vert)$ is a more complicated function related to the function in Fig.~\ref{fig:largeboost} in the case where $\vec p$ is obtained by boosting both a particle at rest and a static spherical bubble.

Because the initial false-vacuum state is not Lorentz invariant, this suppression is consistent with the relativistic invariance of the underlying theory.  We have stressed that we are 
interested in the amplitudes for emission of bubbles with excitations at time zero.  We have seen
that bubbles are mapped by Lorentz transformations into bubbles of different radius, with some linear motion and
some distortion.  As we will explain in Section~\ref{lorentz}, one may also consider the effects of Lorentz transformations
on the particle wave function.  This wave function may be considered as a localized wave packet (amplitudes depend
on the distance from the wall), and the Lorentz transformation of the wave packet changes not only the average momentum but also
the location and shape of the packet.

\subsubsection{Two Analog Models}

The field theory description of these solutions, found in the linearized
approximation, and their energy, momentum, and boosts is straightforward, but 
it is interesting to provide descriptions in terms of small number of degrees of freedom.  Two analogies suggest themselves.

First, for $\Delta R >0$ and no excitations of non-zero $n$ or $\ell$, the bubble has the structure of a particle at rest with {\it negative}
energy.  Thus it is tempting to describe it a negative mass particle, with Lagrangian
\beq
L = - m \sqrt{1-\beta^2};~~~E = {m \over \sqrt{1-v^2}};~~m<0.
\eeq
Small boosts of the particle at rest yield a negative rest energy and a {\it negative kinetic energy}, ${1 \over 2} m \beta^2$.  The corresponding
energy and momentum can be balanced against the energy and momentum of any emitted particle.  In this description, it
is not immediately obvious whether this negative ``kinetic energy" is associated with a suppression of tunneling.

Alternatively, imitating the conventional collective coordinate procedure for solitons, we can introduce collective coordinates for translations,
\beq
\vec x \rightarrow \vec x - \vec x_0(t)
\eeq
with corresponding lagrangian (ignoring, for the moment, mixing with the collective coordinate $R(t)$:
\beq
L = {1 \over 2} 4 \pi S_1 R^2 \dot x^2 -V(R)
\eeq
i.e. a non-relativistic particle of mass $4 \pi S_1 R_0^2$.  This interpretation is consistent with the expression for the 
momentum above; it is, of course, valid only for small velocities.  

In this latter description, Lorentz transformations induce both a change in ${d \vec x \over dt} = \vec v$, and in $\Delta R$.
\beq
\Delta v =-3 {\Delta R \over R_0} \beta
\eeq
In particular, in the limit of small $\beta$ (Galilean transformations), the transformation of the velocity is not that which one has for
ordinary particles; the induced velocity is not the parameter of the transformation,
and there is a change in the radius.  This analogy has other limitations as well.  In particular, the ``mass" in the non-relativistic limit for the collective
mode is not the invariant mass of the configuration.  

Both of these pictures are useful in understanding aspects of this problem, but each has its limitations and they
should be used with some care; it is safest to work directly with the field variables.

\subsection{Lorentz Invariance}
\label{lorentz}

Let us now comment on Lorentz invariance from two points of view.  From the passive viewpoint, Lorentz invariance of the total rate
is straightforward for the $O(3,1)$ symmetric bounce.  We have described a calculation of the tunneling rate
in Minkowski space.  The principle limitation of the calculation is the difficulty of the study of low $\ell$ excitations, but we do not expect this to generate a new divergence.
The calculation involved the study of fluctuations about a Lorentz invariant configuration.  In any other frame, if we define $t=0$ as the time at which the
time derivatives of the background configuration vanish, the calculation is the same.  In other words, rather trivially, as anticipated by Coleman, the calculation
is Lorentz invariant.

From the active viewpoint,  as, for example, used in Sec.~\ref{particleboost}, we need to Lorentz transform states at $t=0$, our emission point.  If we sum over a complete set of states, the Lorentz transformations will just  reorganize the sum.   But there is no reason that the amplitude to produce the transformed field configuration needs to be the same as the amplitude to produce the original configuration.
Indeed, the boost requires knowledge of the time evolution of the states, which will certainly be complicated in the full quantum theory, but in the semiclassical limit we have seen that boosts generate $\ell=0,2$ deformations at $O(\beta^2)$ when acting on larger-than-critical spherical bubbles. We have argued that because of the $\ell=0$ deformations, individual rates are suppressed under boosts, even though the sum over rates is Lorentz invariant. 

We can also note some features of the action of boosts on particle wave packets that might be nucleated along with the bubbles. 
Consider a high-momentum wave packet (semiclassical) located at $t=0$ at a point on or close to the bubble wall.
High-momentum particles are nearly free, so we can consider the effects of Lorentz transformations on the wave packet alone.  For simplicity,
we take the wave to move along the $x$ axis and limit our considerations to Lorentz transformations along this axis.  A wave packet with average momentum $p_0$, group velocity $v_g = \left ( {\partial \omega \over \partial p} \right )_0$, and centered on the wall at the time of
bubble nucleation takes the form
\beq
\delta \phi(x,t) = {\rm exp}\left (-{(x-R_0 - v_g t)^2 \over (\delta x)^2} + i (p_0 x -\omega_0 t) \right ).
\eeq
The Lorentz transform of this configuration behaves at $t=0$ as
\beq
\delta \phi(x,t) = {\rm exp}\left (-{(\gamma x (1-\beta v_g)-R_0 )^2 \over (\delta x)^2} +  i(p_0  \gamma x -\omega_0 \gamma \beta x) \right ).
\eeq 
For large boosts, this corresponds to a wave packet at the nucleation time centered
at
\beq
x = R_0 \sqrt{1-v^2}/(1-v_gv),  
\eeq
i.e. well-inside the bubble for sufficiently large boosts. Therefore, the particle is nucleated further and further away from the increasingly large-radius bubble wall, and so we expect further amplitude suppression with increasing boost parameter (in addition to the previously-discussed suppression induced directly by the growing radial mode.)

From our analysis of this section and the rest of Section~\ref{applications}, we conclude that there is no reason to expect that the Euclidean analysis is in some way misleading.  While the Minkowski
calculation is distinctly more challenging than its Euclidean counterpart, it also leads to results which are finite and Lorentz invariant.

\section{O(4) Non-Invariant Bounces}
\label{nonsymmetric}

It is interesting to consider the behavior of $O(4)$ non-invariant bounces.  Coleman et al. proved that the lowest action
solution is always $O(4)$ invariant.   We will demonstrate below (Section \ref{heuristic}) that
{\it all} thin wall solutions are $O(4)$ invariant.   To the best of our knowledge, non-$O(4)$ invariant solutions have not been exhibited explicitly, but we know of no general argument that non-symmetric solutions do not exist.
 Such solutions continue to Minkowski space solutions which are not $O(3,1)$ invariant, and so
pose the puzzle of Ref.~\cite{dvali} quite sharply:  does one have to integrate separately over all $O(3,1)$ transformations of the solution (and the small fluctuations
about it), and does this yield
  an infinite result?  On the Euclidean side, the  integral over the collective coordinates associated with the
broken $O(4)$ is manifestly finite.  But in continuing to the Minkowski side (if such a continuation can be defined sensibly-- some rotations of a non-$O(4)$-symmetric configuration will break $t\rightarrow-t$ symmetry, leading to imaginary field values under naive analytic continuation), one of the Euclidean angles becomes a boost parameter, and there would appear
to be an infinity of configurations to which one can tunnel.

In the case of $O(4)$ symmetric bubbles and (non-$O(4)$ perturbations around them) we had to deal with the question of what precisely one should
calculate in a  Minkowski space analysis.  We suggested choosing a particular space-like surface (the approximate moment of bubble nucleation) and summing over all possible
configurations on that surface.  We saw that the
Lorentz transforms of such configurations were complicated.  What one might describe as nucleation in one frame with emission of a high
momentum particle on the wall would have a quite different description in another.
This is certainly the case for a classical, non-$O(4)$ invariant solution.  Indeed, there would be no preferred moment of nucleation (in an
arbitrary frame), even in the approximate
sense we considered in the previous section.  If one chooses an arbitrary space-like surface as the nucleation surface, just as in the case of high momentum
excitations we have considered earlier, the amplitude for the appearance of a bubble and the Lorentz transformed bubble need not
be the same on that surface, just as in the case of non-Lorentz invariant excitations of the symmetric bubble.  Of course, lacking explicit solutions, it is hard to sharpen these questions.  But, as in the symmetric case, it would be quite shocking if the
Euclidean computation were not to produce the correct amplitude.

\subsection{Only $O(4)$ Invariant Solutions in the Thin Wall Approximation}
\label{heuristic}

In the thin wall approximation, we can show that the $O(4)$
invariant solution is the only solution that satisfies the equation
of motion with the boundary condition that the field tends to the
false vacuum at infinity.

In this limit, the size of the bubble is much larger than the
thickness of the wall.
Therefore, very close to the wall, any solution will look like a
domain wall solution.
The bubble solution is then characterized by the shape of the wall.
Let $B$ be the interior of the bubble and $\partial B$ its boundary.
We consider only the case where $\partial B$ is topologically a three
sphere $S^3$.
The Euclidean action can be written as
\begin{equation}
S_E = S_1 \int_{\partial B} \sqrt{g} d^3 x - \epsilon \int_B d^4 x,
\end{equation}
where $S_1$ is the tension of the wall, $\epsilon$ the (negative)
energy density of the false vacuum, and $g$ the determinant
of the induced metric of the flat $4$ dimensional metric on $\partial B$.

Note that the action is just the difference between the area of the
bubble and the volume it encloses.
It is well known~\cite{BdE88} that a solution that extremizes this action has
constant positive mean curvature and the sphere is the only embedded
surface that satisfies this condition.
Therefore, we conclude that the sphere $S^3$ is the only type of
solution.
The radius of the sphere can be determined by extremizing the action
for the bubble of radius $r$.
This singles out the Coleman's bubble as the unique solution.

\section{Conclusions}

There is a history of concern about Lorentz invariance in tunneling computations in Minkowski theories, and of divergences arising from Lorentz transformations of particular
tunneling configurations.  To address these issues, we explained why it is crucial to formulate the question of what one actually calculates.  There is no analog of an $S$-matrix for
such problems, and the initial ``state", given that it lives only a finite time, is not itself Lorentz invariant.  For the thin-walled bounce, we argued that it is natural to think in terms
of a bubble nucleation time, and of the complete set of excitations which may be produced at this time.  Quite generally, we demonstrated that production of excitations of energy
$\Delta E$ directly through tunneling (as opposed to through excitation from low lying states after the tunneling event) is suppressed by at least
\beq
e^{-(\Delta E^3/S_1)^{1/2}}
\eeq
(where $S_1$ is the bubble wall tension).  As a result, there is no divergence with high energy states, e.g. with the emission of very high momentum particles or production of
high $\ell$ modes of the bubble.  

We began with the assertion the equivalence of the Minkowski and Euclidean computations usually follows from simple unitarity and analyticity considerations.
While our study has elucidated the structure of the Minkowski computation and the absence of infinities, the actual real-time calculation of the process
is complicated, and we have only dealt with some aspects here.   Indeed, one lesson, if it was not obvious already, is that the Euclidean computation is far simpler
than the Minkowski one.

Ultimately, it is necessary to consider general relativity
in understanding questions of tunneling. Coleman and De Luccia showed there are situations where tunneling does not occur in gravitational theories.  In particular, a zero or small positive
c.c. state will not necessarily tunnel to a state with negative cosmological constant (crunch space-time).  On the other hand, in a suitable limit, gravitational corrections
are unimportant in determining the decay rate.  In flat space theories, we have seen no difficulties
with the conventional analysis.  Indeed, we have seen that the Minkowski version of the computation, while complicated in detail, has a structure which can be readily understood.

Finally, we noted that $O(4)$ non-invariant bounces would be interesting to study from this perspective.  Our analysis of the excitations of the symmetric bubble, however,
indicates that no fundamentally new difficulties are to be expected.  Furthermore, we showed that such asymmetric bubbles do not exist in theories in which the thin wall approximation is applicable.

\subsection*{Acknowledgments} 
MD, PD, and CSP were supported by DOE grant DE-FG02-04ER41286. 

\newpage
\appendix
\section{Large Boosts and Past Bubbles}
As explained in the Introduction, we do not expect the partial decay rates into final states related by a Lorentz boost to be the same. The boost-non-invariance can be understood from the dependence of the rates on multiple length/time scales in the problem. In this appendix we consider one effect of the time scale associated with the intrinsic finite lifetime of the false vacuum.

In the picture of Ref~\cite{dvali}, we consider matrix elements of the form ($\mathcal{J}^{0i}$ is a boost generator)
\begin{align}
\mathcal{M}=\langle d_0|e^{i\omega_{0i}\mathcal{J}^{0i}_t}e^{-i\mathcal{H}(t-t_0)}|i\rangle\;,
\end{align}
where $|d_0\rangle$ is a classically-non-$O(3,1)$-invariant final state and $|i\rangle$ is the false vacuum.  
We can rewrite $\mathcal{M}$ as
\begin{align}
\mathcal{M}=\langle d_0|e^{-i\mathcal{H}(t-t_0)}e^{i\omega_{0i}\mathcal{J}^{0i}_{t_0}}|i\rangle\; ,
\end{align}
which tells us that we can either study the projection of the time evolution of the initial state onto various final states boosted around the event $(t,\vec{0})$, or equivalently we can study the projection onto one final state of the time evolution of various boosts of the initial state around the event $(t_0,\vec{0})$ on the initial data surface. 
We will now analyze the second case.

\begin{figure*}[!t]
\begin{center}
\includegraphics[width=0.50\textwidth]{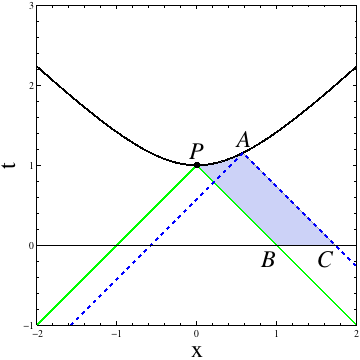} 
\caption{A spacetime diagram showing events (shaded region) that enter the past lightcone (solid green) of the future point $P$ located at (1,0) under a series of Lorentz boosts about the point (0,0). Under one such boost the dashed blue lines are mapped to the solid green lightcone. The shaded region is bounded by the initial data surface (thin black line at $t=0$) in order to show only those points which may contain a nucleation and that can be mapped somewhere into the past of $P$ (including prior to the initial surface) under a boost of magnitude less than or equal to the boost mapping $A$ to $P$.}
\label{spacetime}
\end{center}
\end{figure*}

In particular, since we are interested in decay rates per unit volume, we can ask how large the boost has to be to move a bubble into the past lightcone of a given point in the future of the initial data surface. Once a bubble is present in the past lightcone of the point, the amplitude to nucleate a new bubble at that point goes to zero.
Let us choose our initial surface (where the false vacuum is prepared) at $t=0$ and require that all points below the initial surface are free from bubbles. Our future point is the event $P=(\tau,\vec{0})$; this may be regarded as the farthest point in the future for which it is still probable not to have a bubble in the past lightcone between $P$ and the initial data surface. Under a Lorentz boost about the origin, some events inside the past lightcone are moved out and are replaced by events that were previously outside the lightcone. A two-dimensional illustration is given in Fig.~\ref{spacetime}. 

If the 4-volume of points above $t=0$ that are pushed into the past lightcone by a Lorentz transformation is of the order of the 4-volume of the past lightcone of $P$ above $t=0$, then the boost may be considered the smallest transformation that will with non-negligible probability introduce a bubble somewhere into the past lightcone of $P$ (including prior to $t=0$.) After some geometry, we arrive at the equation
\begin{align}
\frac{\pi\tau^4}{3}=\frac{\pi\tau^4}{6}\left(\frac{v((5-2v)v+11)}{(v-1)^2(v+1)}-3\;\mbox{arcsinh}^{-1} (v\gamma)\right)
\end{align}
where the left-hand side is the 4-volume of the past lightcone and the right-hand side is the 4-volume of points that pass into the lightcone under boosts determined by velocities of magnitude less than or equal to $v$. The right-hand side is obtained by integrating the area of the shaded region in Fig.~\ref{spacetime} over boosts in the other spatial directions.
Since $\tau$ is the only dimensionful parameter in this problem (in the limit of a small, coincident bubble+particle final state), as long as it is finite and nonzero, it drops out of the equation. We can then solve for a constant value of $v$, independent of the lifetime of the system. We find a numerical value of $v_{max}\approx 0.17$.

Consequently, even if all quantum mechanical objections raised in the main text above could be overcome, the contribution to the lifetime from states obtained by boosts greater than $v_{max}$ is likely to be negligible.

\bibliography{vacuumdecay.bib}

\bibliographystyle{unsrt}

\end{document}